\documentclass{elsart}

\usepackage{graphicx}
\usepackage{subfigure}
\usepackage{amssymb}
\usepackage[dvips]{color}

\begin{document}

\begin{frontmatter}

\title{Structure and electronic properties of new model dinitride systems: 
       A density-functional study of $ {\rm \bf CN_2} $, 
       $ {\rm \bf SiN_2} $, and $ {\rm \bf GeN_2} $}

\author{R.\ Weihrich$^{1,2}$}, 
\author{V.\ Eyert$^{1,3}$\corauthref{corauth}},
\corauth[corauth]{Corresponding author. fax: +49 821 598 3262}
\ead{eyert@physik.uni-augsburg.de}
\author{S.\ F.\ Matar$^1$}
\address{$^1$Institut de Chimie de la Mati\`{e}re Condens\'{e}e de Bordeaux, 
         I.C.M.C.B-CNRS, 33608 Pessac Cedex, France, \\
         $^2$Anorganische Chemie, Universit\"at Regensburg, 93040 Regensburg,
         Germany, \\
         $^3$Institut f\"ur Physik, Universit\"at Augsburg, 86135 Augsburg,
         Germany}

\begin{abstract}
The dinitrides $ {\rm CN_2} $, $ {\rm SiN_2} $, and $ {\rm GeN_2} $ in 
assumed pyrite-type structures are studied by means of density functional 
theory using both ultrasoft pseudopotentials and the augmented spherical 
wave (ASW) method. The former two materials constitute the large-$ {\rm x} $ 
limit of the broader class of $ {\rm CN_x} $ and $ {\rm SiN_x} $ compounds, 
which are well known for their interesting mechanical and electronic 
properties. For $ {\rm CN_2} $ a large bulk modulus $ B_0 $ of 405 GPa was 
determined . While $ {\rm SiN_2} $ is found to be a wide band gap compound, 
the calculated gaps of $ {\rm CN_2} $ and $ {\rm GeN_2} $ are considerably 
smaller. The trends in structural and electronic properties as e.g.\ bond 
lengths, band gaps and covalency are well understood in terms of the 
interplay of different types of bonding. 
\end{abstract}

\end{frontmatter}

\section{Introduction}

In recent years nitrogen has turned out to play a key role in todays 
most exciting technological applications of materials science. Important 
examples are the developement of III-V nitrides (AlN, GaN, InN) for 
light emitting diodes (LED) and lasers \cite{nakamura93,okada96}, the 
investigation of $ {\rm Si_3N_4} $ and $ {\rm SiO_xN_y} $ as promising 
candidates for ultrathin high dielectric materials substituting 
$ {\rm SiO_2} $ in the race for ever shrinking field effect transistor 
(FET) gate lengths \cite{konofaos03}, and, finally, new ultra hard 
materials with the proposal of a model carbon nitride $ {\rm C_3N_4} $ as 
the hardest known material \cite{teter96}. However, all attempts to grow 
thin layers of nitrogen rich stoichiometries failed because of the easy 
release of nitrogen due to the high stability of molecular $ {\rm N_2} $ 
\cite{hellgren99}. This major drawback led to propose nitrogen poor 
stoichiometries such as $ {\rm C_{11}N_4} $ \cite{snis99}. Nevertheless, 
the process of $ {\rm N_2} $ release from $ {\rm CN_x} $ has not yet been 
investigated in detail and only recently a preliminary approach based on 
a model structure $ {\rm CN_2} $ was presented \cite{weihrich03}. However, 
our knowledge about this class of nitride systems is far from exhaustive. 
To some extent this is due to the fact that the search for compounds with 
new materials properties is not straightforward and can be experimentally 
very demanding. In this situation, state-of-the-art computer simulations 
built on the principles of quantum mechanics as embodied in density 
functional theory (DFT) are very helpful. In particular, they even allow 
to investigate compounds, which are not yet synthesized. \\
Aiming at a deeper  understanding of ultra hard nitrides we report in the 
present work on investigations of the lightest IV-V compounds 
$ {\rm CN_2} $, $ {\rm SiN_2} $ and $ {\rm GeN_2} $ as crystallizing in 
an assumed cubic pyrite structure \cite{weihrich03}. This choice of 
hypothetical structure is guided by the observation that it allows to 
build up C-N bonds in a three-dimensional network while still satisfying 
the strong N-N bonding. At the same time, the simplicity of the pyrite 
structure offers the advantage that the different types of bondings can 
be easily addressed. Finally, this structure is also adopted by 
other isoelectronic $ {\rm AX_2} $ compounds, as e.g.\ $ {\rm SiP_2} $ 
\cite{chattopadhyay84,brese94}. Further experimental evidence comes from N- 
or C-doped $ {\rm SiP_2} $ and related compounds. In the pyrite structure, 
the N-N entities occupy octahedral holes formed by the C, Si and Ge fcc 
sublattices. The latter are well known from the cubic structures of 
diamond, SiC, SiGe, and TiC. As estimates for amorphous $ {\rm SiN_x} $ 
\cite{martin87} and $ {\rm SiN_2} $ clusters \cite{bae02} show, 
$ {\rm CN_2} $ and $ {\rm SiN_2} $ represent the large-$ {\rm x} $ limit 
of those ultra hard and wide electronic band gap $ {\rm CN_x} $ and 
$ {\rm SiN_x} $ materials, which contain N-N-bonds. \\
From a chemical point of view, the transition from $ {\rm SiP_2} $ 
to $ {\rm SiN_2} $, $ {\rm CN_2} $, and $ {\rm GeN_2} $ involves 
several fundamental issues. These are the behaviour of the light group 
IV compounds C, Si and Ge in octahedral environments as well as that 
of nitrogen in the center of tetrahedra formed by one N and three C (Si/Ge) 
neighbours. The resulting description on the related bonding schemes 
should be found in between Pauling´s covalent bonding \cite{pauling73} 
and ionic concepts like $ {\rm A^{4+}[N_2]^{4-}} $ with the anions 
isoelectronic to $ {\rm [S_2]^{2-}} $ in pyrite $ {\rm FeS_2} $. 
While $ {\rm [N_2]^{2-}} $ was proposed to appear in diacenides 
\cite{auffermann01}, covalent N-N-bonds have been proposed for 
clustered III-V semiconductor nitrides \cite{costales92}. \\
In this work we concentrate, using DFT-based methods, on finding the 
optimal crystal structure parameters of pyrite $ {\rm CN_2} $, 
$ {\rm SiN_2} $ and $ {\rm GeN_2} $. In addition, we aim at relating 
the crystal structure stability to the electronic properties as well as 
to the chemical bonding.

\section{Computational methods}

The electronic structure calculations were performed within the framework 
of density functional theory and the local density approximation using 
two complementary approaches. First, full geometry relaxations including 
optimization of the cell volume and the atomic positions were performed 
with the help of an ultrasoft pseudopotential scheme as implemented in 
the VASP code \cite{kresse93,kresse94,kresse96a,kresse96b}. Calculations 
were converged at an energy cutoff of 434.82 eV for the plane-wave basis 
set. The Brillouin zone integration was performed using 11 to 76 irreducible 
$ {\bf k} $ points and the tetrahedron method \cite{bloechl94}. \\
In a second step, the electronic properties of all three compounds were 
determined using the all-electron augmented spherical wave (ASW) method 
\cite{williams79,eyert00}. In order to represent the correct shape of 
the crystal potential in the large voids, additional augmentation spheres 
were inserted into the open pyrite structure. Optimal augmentation sphere 
positions as well as radii of all spheres were automatically generated 
by the sphere geometry optimization (SGO) algorithm \cite{eyert98b}.
The basis sets comprised $ s $, $ p $, and $ 3d $ states of the group 
IV element, N $ 2s $, $ 2p $ states as well as states of the additional 
augmentation spheres. The Brillouin zone integration was done using an 
increasing number of ${\bf k}$ points within the irreducible wedge ranging 
from 11 to 1135 points, again to ensure convergence of the results with 
respect to the $ {\bf k} $-space grid.

\section{Results and Discussion}

\subsection{Crystal structure optimization}

\begin{figure}[htp]
\centering
\includegraphics[width=0.45\textwidth]{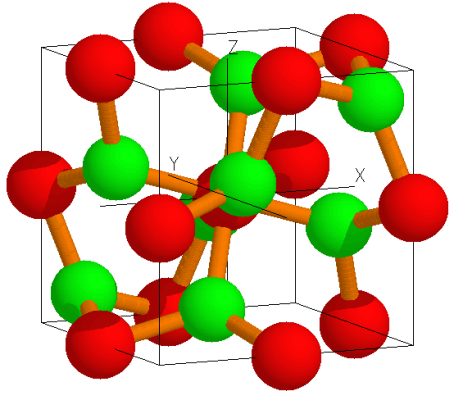}
\caption{Pyrite structure. Group IV element atoms are found at the centers 
         of the cube and the edges, respectively.}
\label{fig:cryst1}
\end{figure}

The pyrite structure $ {\rm AX_2} $ is best described in terms of the 
rocksalt structure with the A atoms and $ {\rm X_2} $ occupying the  
two sublattices such that cubic symmetry with space group {\em Pa$\bar{3}$} 
is preserved \cite{eyert98a}.  As a result, the A atoms are found at the 
center of $ {\rm X_6} $ octahedra, while three A atoms and one X atom 
form distorted tetrahedra about the X sites, see Fig.\ \ref{fig:cryst1}.
The pyrite structure is fully specified by two parameters, namely, the 
lattice constant and the positional parameter $ {\rm x} $ of the X atoms. \\ 
To start with, we performed a structural optimization for $ {\rm SiP_2} $, 
where the theoretical results could be checked against experimental data 
\cite{chattopadhyay84}, see Tab.\ \ref{tab:cryst1}. 
\begin{table}[htp]
\begin{center}
\begin{tabular}{llccccccc}
\hline 
\hline 
                &              & 
    E    &    V   &    a   & $ {\rm x_X} $ & $ {\rm d_{X-X}} $ & 
 $ {\rm d_{A-X}} $ & $ {\rm \frac{d_{X-X}}{d_{A-X}} } $ \\
\hline 
$ {\rm SiP_2} $ & exp & 
         &        & 5.7060 & 0.3906 & 2.162 & 2.397 & 0,902 \\ 
\hline 
$ {\rm SiP_2} $ & LDA &
  -6.213 & 180.85 & 5.6551 & 0.3905 & 2.145 & 2.376 & 0,903 \\
\hline 
$ {\rm SiN_2} $ & LDA          &
  -8.806 &  86.32 & 4.4195 & 0.4050 & 1.454 & 1.886 & 0,771 \\
\hline 
$ {\rm  CN_2} $ & LDA          &
  -7.948 &  63.13 & 3.9818 & 0.4029 & 1.339 & 1.695 & 0,790 \\
\hline 
$ {\rm GeN_2} $ & LDA          &
  -7.785 & 100.37 & 4.6474 & 0.4113 & 1.428 & 1.998 & 0,715 \\
\hline 
\hline 
\end{tabular} 
\end{center}
\caption{Crystallographic data of the optimized structures. 
         Lengths in \AA \ and energies in eV/atom. 
         Experimental data taken from Ref.\ \protect \cite{chattopadhyay84}.} 
\label{tab:cryst1}
\end{table}
As expected, the lattice constant is slightly underestimated by the LDA 
calculations. In contrast, the internal parameter $ {\rm x_P} $ is almost 
exactly reproduced and so is the fraction $ {\rm \frac{d_{X-X}}{d_{A-X}} } $, 
with the P-P distance being shorter than the Si-P bond. \\
On going to the nitride we observe a drastic volume decrease with a 
calculated lattice constant of $ \approx 4.42 $ \AA \ close to the value 
of 4.348 \AA \ measured for $ {\rm SiC} $ \cite{pearson91}. In addition, 
the distance fraction $ {\rm \frac{d_{X-X}}{d_{A-X}} } $ has considerably  
decreased. While the N-N bond length amounts to 1.454 \AA, the Si-N distance 
of 1.886 \AA \ is very similar to the corresponding value of 1.889 \AA \ 
observed for $ {\rm Si_3N_4} $ and is indicative of a single bond. 
For $ {\rm GeN_2} $, our calculations yield a bigger cell, albeit, with 
a still shorter N-N distance of 1.428 \AA. \\
Finally, optimizing the pyrite structure for $ {\rm CN_2} $, we obtain an  
even smaller lattice constant of 3.982 \AA. Together with the calculated 
internal parameter of $ {\rm x_N = 0.403} $ this leads to an N-N distance 
of 1.34 \AA. While still being much larger than in molecular $ {\rm N_2} $ 
(1.09 \AA), this value is considerably smaller than in $ {\rm SiN_2} $ 
and $ {\rm GeN_2} $ strengthening the double bond. At the same time the 
C-N distance of 1.695 \AA \ is significantly larger than the 1.47 \AA \   
measured for the C-N single bonds of $ {\rm C_3N_4} $ indicating an even 
lower order. Applying the usual estimates for the hardness of 
$ {\rm CN_x} $ compounds \cite{mattesini00,mattesini02} we calculate 
a bulk modulus of $ B_0 = 405 $ GPa from a Birch fit \cite{birch78}. 
Since this is harder than cubic BN,  $ {\rm CN_2} $ in the pyrite type 
structure may be regarded as a high-density ultra hard material. Since 
the energy values given in Tab.\ \ref{tab:cryst1} are relative to the 
atomic energies and thus not indicative for the stability of the compounds,  
we performed additional supercell calculations for atomic C (Si, Ge) as 
well as molecular $ {\rm N_2} $ and obtained a heat of formation of 
-4 eV/atom.

\subsection{Electronic structure}

\begin{figure}[htp]
\centering
\subfigure{\includegraphics[width=0.48\textwidth]{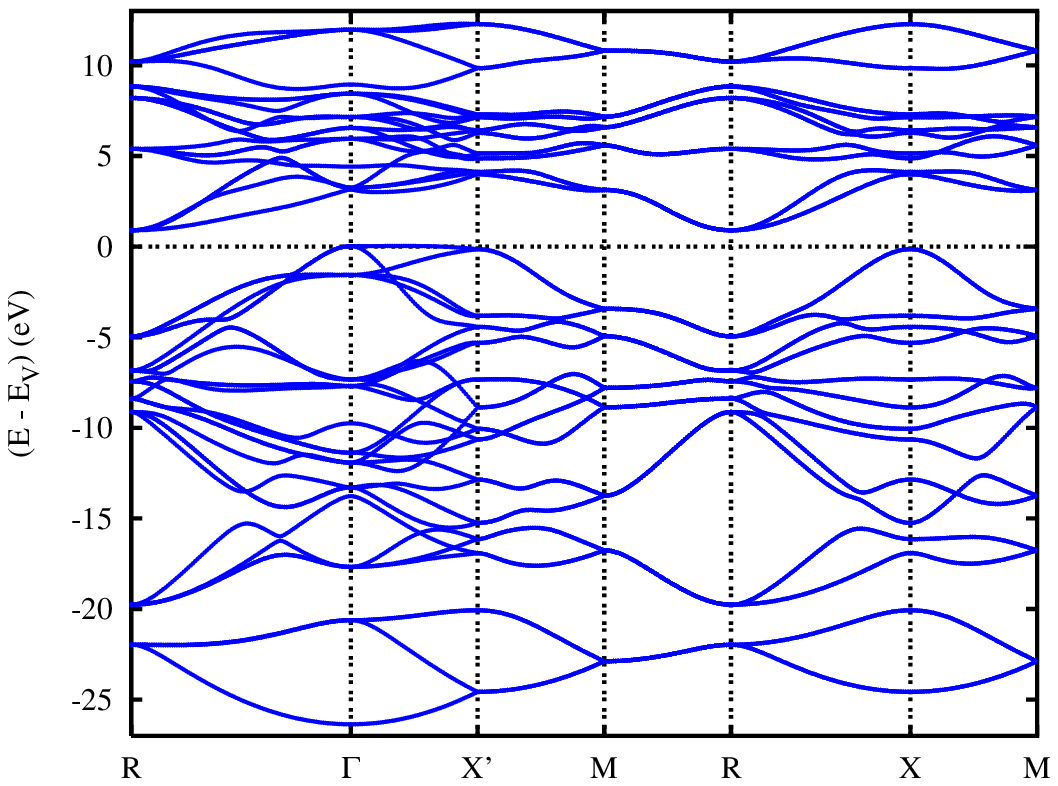}}
\hspace{0.01\textwidth}
\subfigure{\includegraphics[width=0.48\textwidth]{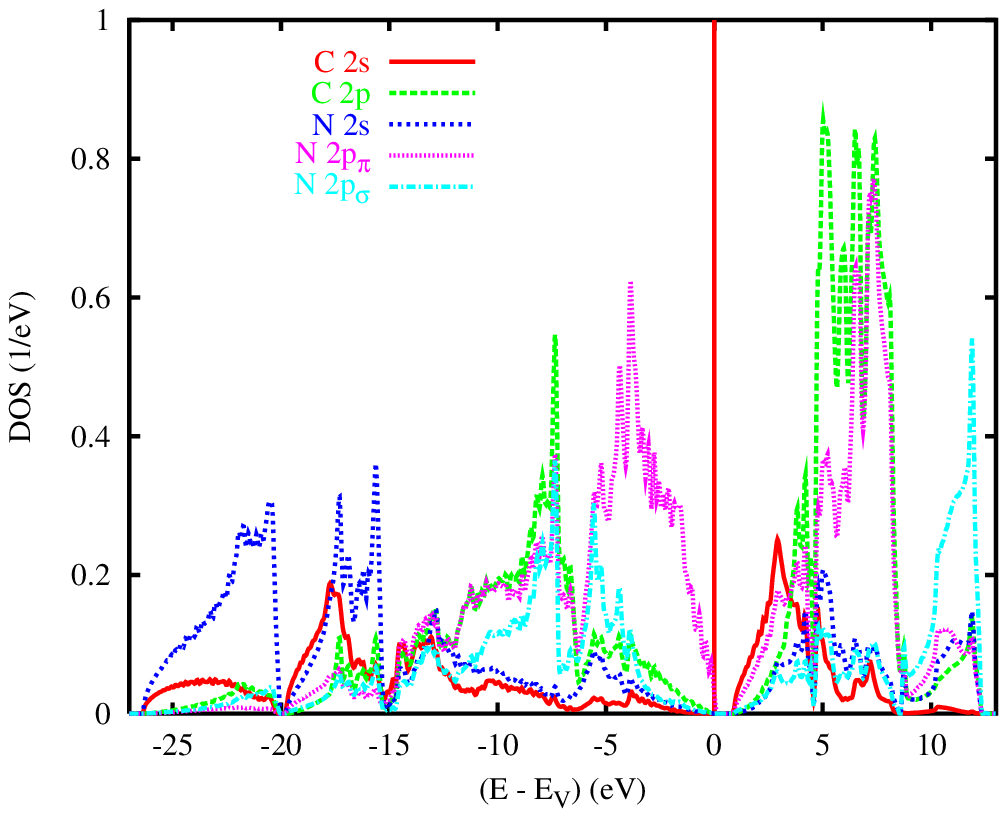}}
\caption{Electronic structure and partial densities of states (DOS) of 
         $ {\rm CN_2} $.}
\label{fig:dos1}
\end{figure}

\begin{figure}[htp]
\centering
\subfigure{\includegraphics[width=0.48\textwidth]{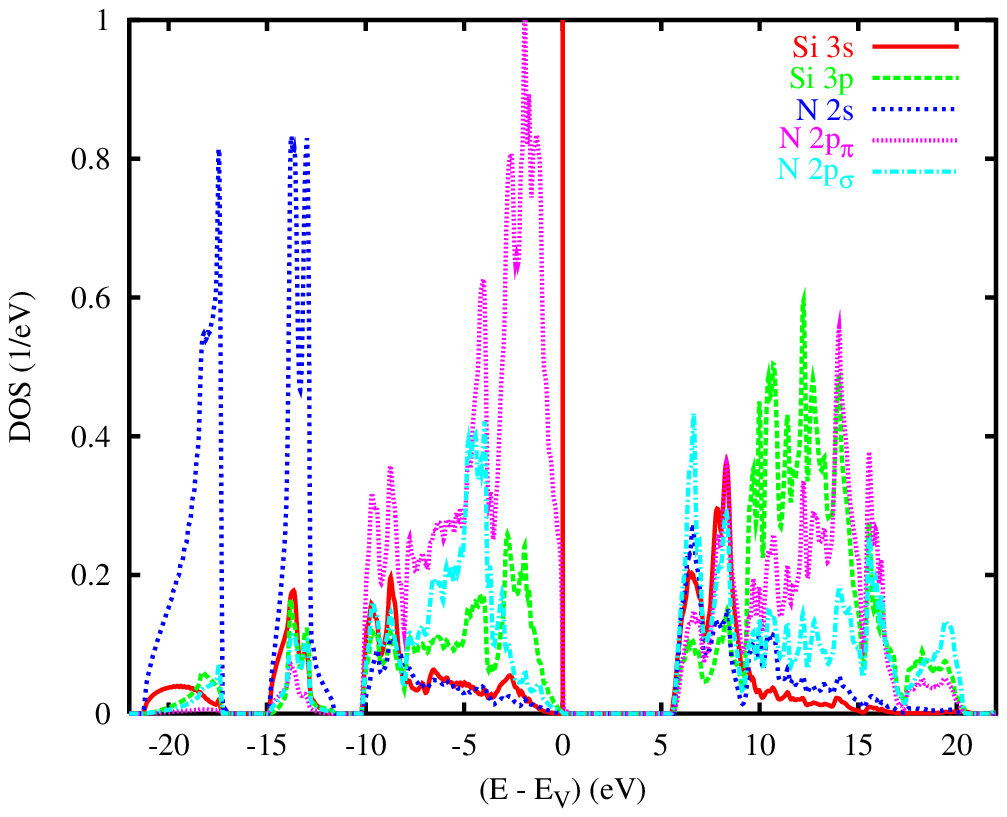}}
\hspace{0.01\textwidth}
\subfigure{\includegraphics[width=0.48\textwidth]{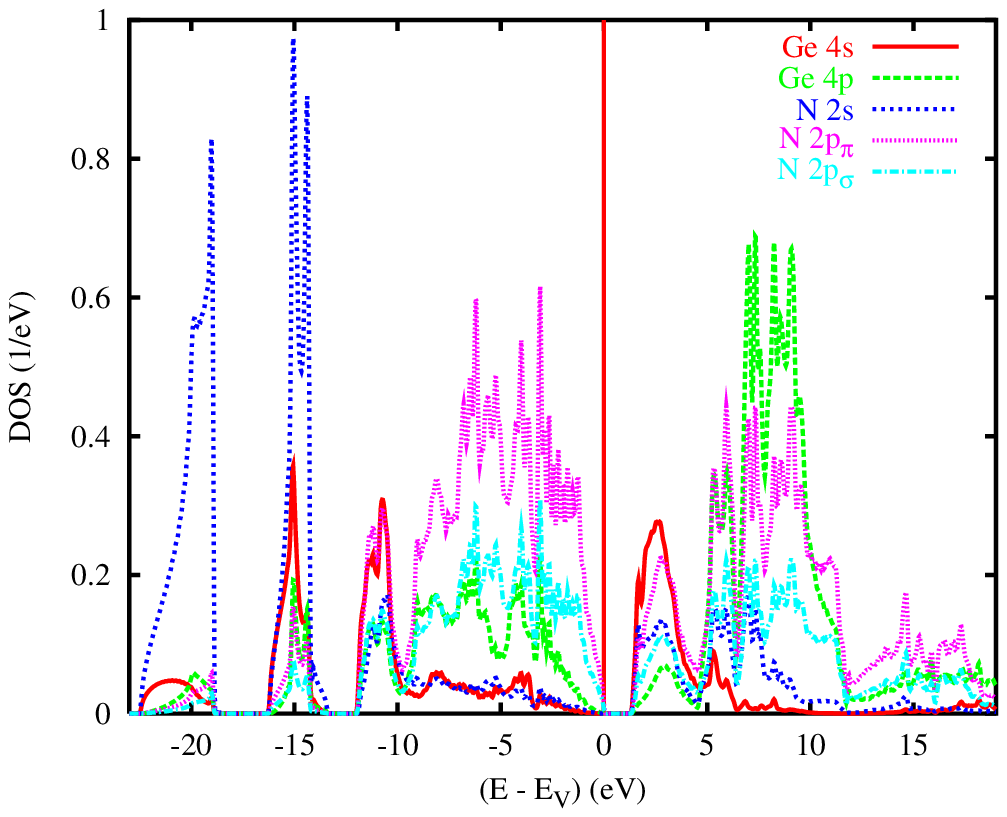}}
\caption{Partial densities of states (DOS) of $ {\rm SiN_2} $ and 
         $ {\rm GeN_2} $.}
\label{fig:dos2}
\end{figure}

The band structure of $ {\rm CN_2} $ along selected high symmetry lines 
within the first Brillouin zone of the simple cubic lattice 
\cite{eyert98a} as well as the partial densities of states (DOS) of all 
three compounds are shown in Figs.\ \ref{fig:dos1} 
and \ref{fig:dos2}.
In these figures four groups of bands are identified. The two low-energy 
groups of bands each comprise four single bands. Bands are most easily 
counted along the line M-R, where they are fourfold degenerate. According 
to the partial densities of states these bands trace back mainly to the 
N $ 2s $ states, which form a bonding and antibonding 
subset of several eV width. Whereas the large splitting between these 
two groups is due to the short N-N distance the intrinsic band widths 
of both groups mainly result from the dispersion across the 
face-centered cubic (fcc) lattice formed by the nitrogen pairs. 
This is concluded from the level sequence at the $ {\rm \Gamma} $ 
point into a lower nondegenerate and an upper threefold degenerate 
state. The latter results from folding the fcc bands into the simple 
cubic Brillouin zone (for the antibonding group of bands the sequence 
of levels is reversed). Further support comes from the striking 
similarity of the density of states especially of the bonding bands 
to that of a band calculated within the tight-binding approximation 
for an fcc lattice. Band widths are largest for $ {\rm CN_2} $, which 
has the smallest lattice constant. \\
The electronic structure of the valence and conduction band are influenced 
by different types of bondings. Both groups comprise 20 bands, which derive 
from hybridization of all orbitals except for the N $ 2s $ states. Note 
that we have used a rotated coordinate system to separate the N $ 2p $ 
partial densities of states into contributions from orbitals mediating 
$ \sigma $- and $ \pi $-type bonding within the $ {\rm N_2} $ pairs. The 
calculated optical band gaps are 0.9, 5.5, and 1.4 eV for $ {\rm CN_2} $, 
$ {\rm SiN_2} $ and $ {\rm GeN_2} $, respectively. \\
Common to all three nitrides is the strong bonding between group IV element 
$ p $ states and those of nitrogen, which leads to bonding and antibonding 
bands in the energy regions [-12:-6] and [5:9] eV for $ {\rm CN_2} $, 
[-10:-4] and [10:15] eV for $ {\rm SiN_2} $, and [-9:0] and [5:18] eV for 
$ {\rm GeN_2} $, respectively. Since the nitrogen atoms are coordinated by 
three group IV atoms this bonding is rather isotropic and, hence, involves 
both the $ 2p_{\sigma} $ and $ 2p_{\pi} $ states in the same manner with 
the ratio of 1:2 of their partial DOS reflecting their respective weight. 
Except for the carbon compound the unoccupied antibonding bands are 
dominated by the group IV element $ p $ states, whereas the N $ 2p $ 
orbitals show a larger contribution to the bonding bands. \\
Differences between the three compounds, which cause, in particular, the 
spread in calculated optical band gaps, can be traced back mainly to the 
different influence of two additional types of bonding. These are, 
respectively, the bonding between the group IV element $ s $ and the 
N $ 2sp $ states as well as the strong $ \sigma $- and $ \pi $-type overlap 
within the nitrogen pairs. The former leads to considerable contributions 
of the C/Si/Ge $ s $ states to the second group of bands as well as to the 
lower edges of the valence and conduction bands. Across the series, 
$ s $-$ p $ hybridization of the group IV element decreases and, on going 
from C to Ge, the bonding and antibonding bands involving the $ s $ states 
are lowered in energy relative to the bands derived from the group IV element 
$ p $ states. This effect is already visible for $ {\rm SiN_2} $ and becomes 
most obvious for $ {\rm GeN_2} $, where it leads to the substantial decrease 
of the optical band gap. \\ 
To the contrary, the $ \sigma $- and $ \pi $-type overlap within the 
$ {\rm N_2} $ pairs causes splitting of the $ 2p_{\sigma} $ and $ 2p_{\pi} $ 
states into bonding and antibonding bands. In the partial DOS, those N $ 2p $ 
states, which are subject to overlap within the pairs, can be easily 
identified by the smaller C/Si/Ge contributions in the respective energy 
range. However, due to the large N-N distance in both $ {\rm SiN_2} $ and 
$ {\rm GeN_2} $, this bonding/antibonding splitting is rather weak in these 
two nitrides. This is contrasted by the situation in $ {\rm CN_2} $, where 
the close coupling of the nitrogen atoms leads to strong splitting into 
bonding and antibonding bands showing up in the energy intervals from 
[-6:0] and [8:12] eV, respectively. To conclude, the N $ 2p $ states are 
subject to two different types of bonding, namely, that inside the pairs 
and that with the group IV element $ sp $ states. The bonding and 
antibonding states resulting from both types of chemical bonds end up in 
different energy regions. In particular, for $ {\rm CN_2} $ the bonding 
bands resulting from N-N $ 2p $ overlap are higher in energy than the bands 
growing out of C/Si/Ge $ sp $-N $ 2p $ overlap. As a consequence, the 
optical band gap is considerably reduced as compared to $ {\rm SiN_2} $.

\section{Conclusions}

First principles calculations for the dinitrides $ {\rm CN_2} $, 
$ {\rm SiN_2} $ and $ {\rm GeN_2} $ in assumed pyrite-type structures 
result in stable crystals. A bulk modulus of $ B_0 $ = 405 eV is 
calculated for $ {\rm CN_2} $, which thus can be regarded as an ultra hard 
material. The electronic structures of all three nitrides are governed by 
three different types of bonding. Overlap of the C/Si/Ge $ s $ and $ p $ 
orbitals with the N $ 2p $ states lead to the formation of bonding and 
antibonding bands, which give rise to the valence and conduction groups of 
bands, respectively. In both groups, the C/Si/Ge $ s $ derived states are 
found at the lower band edge. Lowering of the $ s $ states relative to 
the $ p $ states due to decreasing $ s $-$ p $ hybridization across the 
series C-Si-Ge results in a depression of the optical band gap of 
$ {\rm GeN_2} $ as compared to that of $ {\rm SiN_2} $. Close coupling 
within the $ {\rm N_2} $ pairs especially in $ {\rm CN_2} $ gives rise to 
a third type of bonding, namely $ \sigma $- and $ \pi $-type overlap of 
the N $ 2p $ states. The resulting bonding and antibonding bands are 
found at the upper edges of the valence and conduction bands of 
$ {\rm CN_2} $ and again lead to a strong decrease of the optical band 
gap as compared to $ {\rm SiN_2} $.

\section*{Acknowledgements}
We are grateful to Profs.\ K.-J.\ Range and J. Etourneau for a lot of 
fruitful discussions, continuous interest, and support of this work.  
R.\ W.\ and V.\ E.\ gratefully acknowledge the kind hospitality of the 
ICMCB and the University Bordeaux 1. 
R.\ W.\ acknowledges financial support by the Training and Mobility Network 
"New carbon based ultra hard Materials", 
1997-2002 "FMRX-CT97-0103 (DG 12-MSPS)". 
This work was supported by the Deutsche Forschungsgemeinschaft through 
SFB 484, Augsburg. 
Calculations were done on the Regatta IBM P690 of the M3PEC intensive 
numerical computations facility of the University Bordeaux 1.

\end{document}